\documentclass[amsmath,superscriptaddress,showpacs,prb,twocolumn]{revtex4-1}
\usepackage{bbm}
\usepackage{bm}
\usepackage{enumerate}
\usepackage{graphicx}
\usepackage[dvips]{epsfig}
\usepackage{epsf}
\usepackage[latin1]{inputenc}
\usepackage{amsmath}
\usepackage{amsfonts}
\usepackage{amssymb}
\usepackage{xcolor}

\bibpunct{[}{]}{;}{n}{}{}

\begin{document}
	\title{Spin superfluidity in noncollinear antiferromagnets}
	
	\author{Bo Li}
\affiliation{Department of Physics and Astronomy and Nebraska Center for Materials and Nanoscience, University of Nebraska, Lincoln, Nebraska 68588, USA}

\author{Alexey A. Kovalev}
\affiliation{Department of Physics and Astronomy and Nebraska Center for Materials and Nanoscience, University of Nebraska, Lincoln, Nebraska 68588, USA}
	
\date{\today}

\begin{abstract}
We explore the spin superfluid transport in exchange interaction dominated three-sublattice antiferromagnets. The system in the long-wavelength regime is described by an $SO(3)$ invariant field theory. Additional corrections from  Dzyaloshinskii-Moriya interactions or anisotropies can break the symmetry; however, the system still approximately holds  a $U(1)$-rotation symmetry. Thus, the power-law spatial decay signature of spin superfluidity is identified in a nonlocal-measurement setup where the spin injection is described by the generalized spin-mixing conductance.  We suggest iron jarosites as promising material candidates for realizing our proposal.
\end{abstract}

\maketitle

Spintronics has been extremely successful in combining advanced theoretical concepts with practical applications and experiment \cite{spintronics}.  
Materials with antiferromagnetic ordering are a focus of active research in spintronics due to many desirable properties such as spin dynamics in terahertz range \cite{Olejnk2018}, the absence of stray fields, and insensitivity to the presence of magnetic fields \cite{RevModPhys.90.015005}. In addition, antiferromagnetic insulators are characterized by long spin diffusion length associated with transport of magnons, making them particularly suitable for spintronics applications such as  low dissipation electronic devices \cite{Lebrun2018}. 

Magnetic insulators can also transport spins in a regime in which the transport can be describe as spin superfluidity \cite{Sonin2010,PhysRev.188.898}. In easy-plane magnets, the spin is then transported over large distances by the coherent order parameter precession \cite{PhysRevLett.112.227201,PhysRevB.90.094408,PhysRevLett.115.237201}. The power-law decay of spin current can enable spin transport over longer distances compared to the diffusive regime \cite{PhysRevLett.87.187202,PhysRevLett.112.227201,PhysRevB.95.144432,PhysRevB.90.094408,PhysRevLett.116.117201,PhysRevLett.115.237201,PhysRevB.96.134434,Yuan2018,Stepanov2018}. Nevertheless, in ferromagnets the dipole interaction can limit the range of spin superfluid transport \cite{PhysRevLett.115.237201}. On the other hand, collinear antiferromagnetic insulators could provide a viable platform for realizing the spin superfluidity \cite{PhysRevB.90.094408,PhysRevLett.118.137201,PhysRevLett.116.216801} as demonstrated in experiments on Cr$_2$O$_3$ \cite{Yuan2018} and antiferromagnetic $\nu=0$ quantum Hall state of graphene \cite{Stepanov2018}. 

Noncollinear antiferromagnets (nAFM) are yet another viable platform for realizing spin flows \cite{PhysRevB.100.100401,PhysRevB.99.224410,PhysRevB.101.035104}. The noncollinear conducting magnets can exhibit a multitude of phenomena associated with topology of electronic bands \cite{mejkal2018}, e.g., Mn$_3$X (X = Ge, Sn, Ga, Ir, Rh, or Pt) magnets exhibit the anomalous \cite{PhysRevLett.112.017205} and spin \cite{PhysRevLett.119.187204} Hall responses. Various magnon-mediated responses relying on magnon spin-momentum locking, topology of magnonic bands, and coupling to phonons have been studied theoretically, promising observation of spin related phenomena in insulating antiferromagnets \cite{PhysRevB.95.014422,PhysRevB.98.094419,PhysRevB.99.014427,PhysRevB.100.064412,PhysRevLett.119.107205,PhysRevB.101.024427,PhysRevResearch.2.013079,Park2020,PhysRevLett.123.237207}.

In this work, we analytically study viability of spin superfluid transport in insulating nAFM. In general, the U(1) symmetry of magnetic ordering can be hampered by various anisotropies. The highly symmetric hexagonal environment considered in this work can be beneficial for realizing spin superfluid transport. Hexagonal nAFM can exhibit relevant phenomena, e.g., the appearance of domain walls \cite{PhysRevB.93.134429,PhysRevB.100.054415} and Goldstone modes \cite{2020arXiv200402790D}. Furthermore, spin superfluid transport has been studied numerically in a triangular nAFM \cite{2020arXiv200513481G}. 
In this work, we offer analytical results with a detailed discussion of the generalized spin-mixing conductance and spin current injection into nAFM. We identify the power-law decay feature of the spin superfluid transport in a nonlocal experimental setup. Our simple results can help in designing and interpreting experiments on spin superfluidity in nAFM.

\textit{Long-wavelength Hamiltonian}---
In nAFM, the exchange interaction is often dominant, which approximately endows the system with an $SO(3)$ symmetry given that all other interactions, e.g., anisotropy, DMI, are very weak. We start with constructing a long-wavelength $SO(3)$ field theory to describe the nAFMs and regard other weak terms as additional perturbations. In a two-dimensional nAFM with three sublattices (e.g., kagome, triangular), the exchange interactions favor fully compensated spin configurations, which in the presence of other interactions may acquire a very small net magnetization. Therefore, we parametrize the spins $\boldsymbol S_i$ of length $S$ in each triangular plaquette as \cite{PhysRevB.39.6797}
\begin{eqnarray}\label{Spinfield}
\boldsymbol S_i=S\hat R(\boldsymbol n_i+ \boldsymbol L)/(1+2\boldsymbol L\cdot\boldsymbol n_i+L^2)^{1/2},
\end{eqnarray}
where $\boldsymbol n_i$ ($i=1,2,3$) sets a reference ordered state allowed by exchange interactions with
\begin{eqnarray}\label{eq:reference}
&&\boldsymbol n_1=(0,1,0),\quad \boldsymbol n_2=(-\frac{\sqrt{3}}{2},-\frac{1}{2},0),\nonumber\\
&&\boldsymbol n_3=(\frac{\sqrt{3}}{2},-\frac{1}{2},0);
\end{eqnarray}
$\hat R\in SO(3)$ is a rotation matrix which generates degenerate states by acting on the reference state; $\boldsymbol L$ describes small deviation from the compensated spin structure  with the magnitude $L\ll 1$. $\hat R$ and $\boldsymbol L$ together generate all possible spin configurations on three sublattices.   To the leading order in $\boldsymbol L$, $\boldsymbol S_i=S\hat R[\boldsymbol n_i+\boldsymbol L-\boldsymbol n_i(\boldsymbol L\cdot \boldsymbol n_i)]$, and the net angular momentum density,
\begin{eqnarray}
\boldsymbol m=\hbar/A_{\text{uc}}\sum_i\boldsymbol S_i=3 s\hat R(\hat T\boldsymbol L),
\end{eqnarray}
where $s=S \hbar/A_{\text{uc}}$, $T_{ab}=\delta_{ab}-\frac{1}{3}\sum_in_{i}^an^b_{i}$, and $A_{\text{uc}}$ is the area of a unit cell.\\

With the forgoing parametrization, the system is generally described by a Lagrangian \cite{PhysRevB.39.6797},
\begin{eqnarray}\label{Lagrangian}
\mathcal L=\boldsymbol m\cdot \boldsymbol{\Omega}_t-\frac{\boldsymbol m^2}{2\varrho}-\text{tr}[\hat P_{ij}\partial_i\hat R^T\partial_j \hat R]
\end{eqnarray}
where $(\Omega_t)_{i}=-\frac{1}{2}R_{i\alpha}\epsilon_{\alpha\beta\gamma}(\hat R^T\partial_t\hat R)_{\beta\gamma}$. Here, the first term is derived from the spin kinetic energy; the second term originates from the exchange interaction, e.g., for the nearest exchange $J$, we obtain $\varrho\propto \hbar^2/(JA_{\text{uc}})$; the last term describes the second-order gradient expansion of exchange coupling with tensor $\hat P_{ij}$ encoding the exchange interactions and lattice geometry. From the Euler-Lagrange equation~\cite{PhysRevB.93.134429}, we obtain: 
\begin{eqnarray}\label{Totalspin}
\boldsymbol m=\varrho \boldsymbol{\Omega}_t,
\end{eqnarray}
from which the field $\boldsymbol m$ can be removed from Eq.~\eqref{Lagrangian}, i.e.,
\begin{eqnarray}\label{Sigmamodel}
\mathcal L=\frac{\varrho}{4}\text{tr}[\partial_t\hat R^T\partial_t \hat R]-\text{tr}[\hat P_{ij}\partial_i\hat R^T\partial_j \hat R].
\end{eqnarray}
This is the so-called nonlinear $\sigma$ model \cite{PhysRevB.39.6797,PhysRevLett.68.1762}.

\begin{figure}
	 \includegraphics[width=0.9\linewidth]{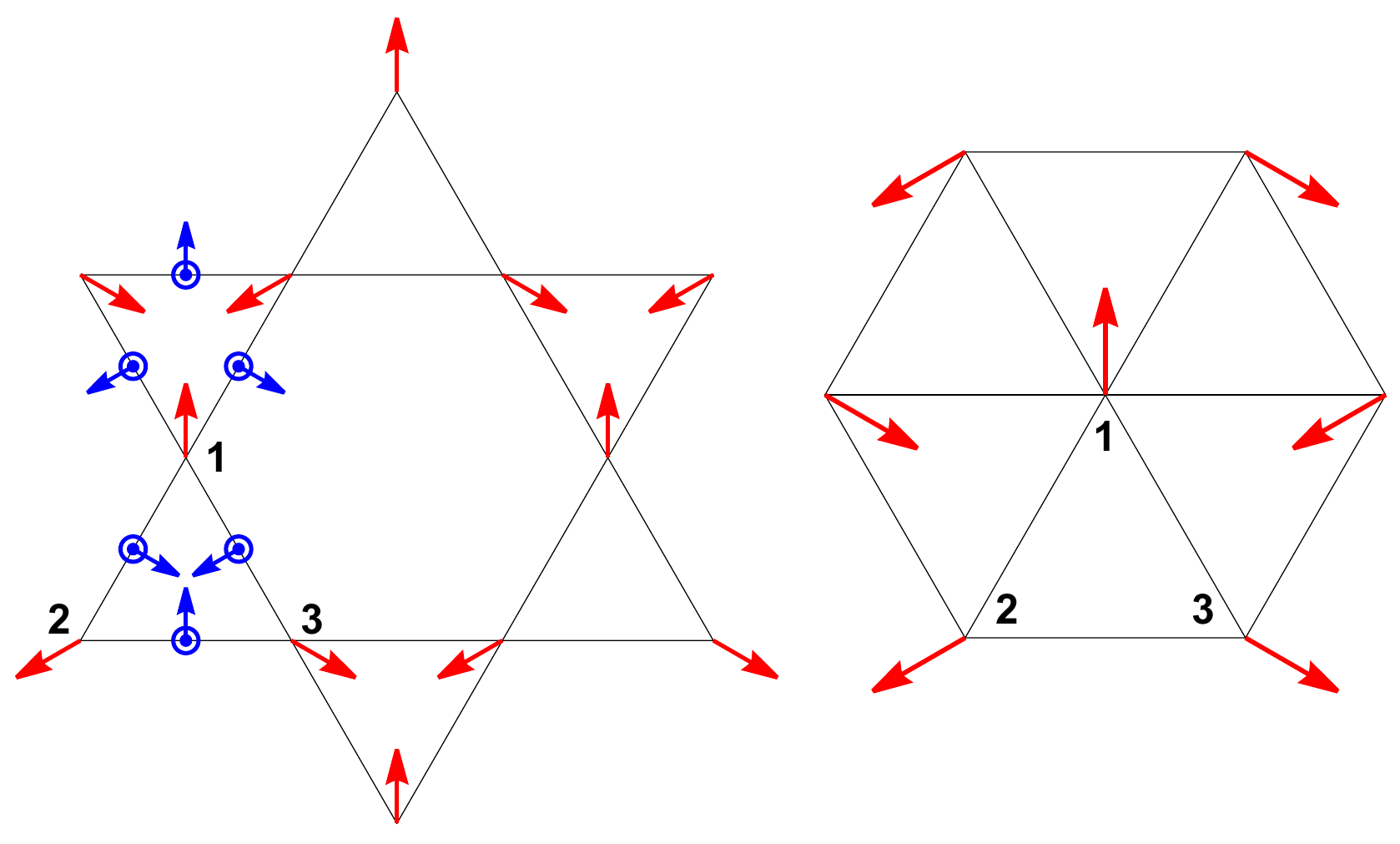}
 \caption{Noncollinear kagome (left) and triangular (right) antiferromagnets. The red arrows indicate the spin directions of the reference state. The blue arrows (left) indicate the out-of-plane and in-plane DMI vectors; the in-plane DMI vectors are defined for anti-clockwise direction in each triangular plaquette. }\label{fig_Symmetry}
\end{figure}

To determine the tensor $\hat P_{ij}$ for a hexagonal-symmetry lattice, kagome or triangular, when exchange is the dominant interaction,  we explore the spin wave behavior by following Ref. \cite{2020arXiv200402790D}. We consider small fluctuations upon the reference state by using   $\hat R=\exp[-i\boldsymbol{\theta}\cdot \hat{\boldsymbol{\mathcal J}}]$, where $(\mathcal J_i)_{jk}=-i\epsilon_{ijk}$ with $\epsilon_{ijk}$ being the Levi-Civita tensor and a vector $\boldsymbol\theta$ describes the small deviation with $|\boldsymbol\theta|\ll 1$. The spin-wave energy is obtained from the leading-order expansion of the second term in Eq.~\eqref{Sigmamodel}, 
\begin{eqnarray}\label{eq:SW}
\mathcal U&&=\text{tr}[\hat P_{ij}\hat{\mathcal J}_k\hat{\mathcal J}_l]\partial_i\theta_k\partial_j\theta_l\nonumber\\
&&= \partial_i\theta_k\partial_j\theta_k\text{tr}[\hat P_{ij}]-P_{ij,kl}\partial_i\theta_k\partial_j\theta_l.
\end{eqnarray}
Due to the highly symmetric hexagonal environment, $\theta_z$ acts as a scalar and $\theta_x,\theta_y$ act as two components of a vector under symmetry transformations.
Using the symmetry constraints, we recover the form of $\hat P_{ij}$ tensor:
\begin{eqnarray}\label{eq:tensors}
&&P_{ij,kl}=\eta\delta_{ij}\delta_{kl}+\lambda\delta_{ik}\delta_{jl}+\mu\delta_{il}\delta_{jk}\quad \text{for}\quad k,l=1,2,\nonumber\\
&&P_{ij,zz}=\kappa\delta_{ij}.
\end{eqnarray}
where the arbitrary coefficients $\kappa$, $\eta$, $\lambda$, and $\mu$ scale as the exchange strength $J$. For triangular and kagome lattices with the nearest exchange interaction, their values are summarized in Table.~\ref{table_coefficient}. 

From Eq.~\eqref{Sigmamodel}, the spin-wave Lagrangian reads $\mathcal L=(\varrho/2)(\dot{\boldsymbol{\theta}})^2-\mathcal U$. By using Eq.~\eqref{eq:tensors}, we can obtain  three linearly dispersive Goldstone modes \cite{2020arXiv200402790D} $\omega_i=v_ik$ with
$v_1=\sqrt{2(2\eta+\lambda+\mu)/\varrho}$,
$v_2=\sqrt{2(\kappa+\eta+\lambda+\mu)/\varrho}$,
$v_3=\sqrt{2(\kappa+\eta)/\varrho}$.
Here, $\omega_1$ corresponds to the scalar mode $\theta_z$, and $\omega_{2,3}$ comes from the vector modes $\theta_x,\theta_y$ \cite{2020arXiv200402790D}.

\begin{table}[ht]
	\centering
	\begin{tabular}{|c| c| c| c c c c|}
		\hline\hline 
	 Lattice& $\varrho$ &$A_{\text{uc}}$& $\kappa$ & $\eta$&$\lambda$&$\mu$ \\ [0.5ex] 
		\hline 
	Triangular& $\frac{2\hbar ^2}{9\sqrt{3}Ja^2}$&$\frac{3\sqrt{3}a^2}{2}$&0&$\frac{\sqrt{3}JS^2}{8}$&0&0\\
	Kagome&$\frac{\hbar^2}{4\sqrt{3}Ja^2}$&$2\sqrt{3}a^2$&0& 0&$\frac{\sqrt{3}JS^2}{16}$&$\frac{\sqrt{3}JS^2}{16}$
		\\[1ex]
		\hline 
	\end{tabular}
	\caption{Parameters describing spin-wave excitations in triangular and kagome lattices with only the nearest exchange interaction. Here, $a$ is the lattice constant.} \label{table_coefficient}
\end{table}

\textit{DMI and anisotropy}---
In nAFMs, the field theory used to describe exchange interactions should be modified by the aforementioned weak interactions that remove the $SO(3)$ symmetry and gap out the Goldstone modes. To take these interactions into account in centrosymmetric crystals, we first consider their microscopic expressions.

In a kagome lattice (see Fig.~\ref{fig_Symmetry}), we consider the DMI term, e.g. typical to jarosites~\cite{PhysRevB.66.014422},
\begin{eqnarray}\label{eq:kagomeDMI1}
H_{D}=\sum_{i,j}\boldsymbol D_{ij}\cdot(\boldsymbol S_i\times\boldsymbol S_j),
\end{eqnarray}
where $\boldsymbol D_{ij}=D_{z}\hat{\boldsymbol z}+\boldsymbol D_{\parallel}$, $\boldsymbol D_{\parallel}=D_p\hat{\boldsymbol n}_{ij}$ with $\hat{\boldsymbol n}_{12}=(\frac{\sqrt{3}}{2},-\frac{1}{2},0)$, $\hat{\boldsymbol n}_{23}=(0,1,0)$, and $\hat{\boldsymbol n}_{31}=(-\frac{\sqrt{3}}{2},-\frac{1}{2},0)$. We first consider DMI to the leading-order in spatial gradients and obtain the energy density,
\begin{eqnarray}\label{KagomeDMI}
\mathcal H_{D}
\approx-i\text{tr}[\hat{\boldsymbol X}\cdot(\hat R^T\hat{\boldsymbol{\mathcal J}}\hat R)],
\end{eqnarray}
where $(\hat X_k)_{ab}=\sum_{i,j=1}^3S^2 \mathcal D^k_{ij}n_i^an_j^b$ and $\mathcal D^k_{ij}= D^k_{ij}/A_{\text{uc}}$. By using Eq.~\eqref{KagomeDMI} and the representation of the rotation matrix, $\hat R=\exp[-i\boldsymbol\theta\cdot\hat{\boldsymbol{\mathcal J}}]\exp[-i \phi \hat{\mathcal J}_z]$, the leading correction of the DMI term is obtained by expanding Eq.~\eqref{KagomeDMI} to the lowest order of $\theta_i$ ($i=x,y,z$),
\begin{eqnarray}
\mathcal \delta \mathcal U=\frac{\Delta}{2}(\theta_x^2+\theta_y^2),
\end{eqnarray}
where $\Delta=-3\sqrt{3}\mathcal D_zS^2>0$. The out-of-plane DMI suppresses spin rotations other than those with respect to $z$-axis, reducing the SO(3) symmetry to a $U(1)$ rotation symmetry. 
Furthermore, the DMI in Eq.~\eqref{eq:kagomeDMI1} constrains the ground state of the system and gaps out the Goldstone modes $\omega_{i}=\sqrt{v_i^2k^2+\Delta/\varrho}$ for $i=2,3$, while the $\omega_1$ mode is intact.

To capture a small gap in the $\omega_1$ mode, we expand  Eq.~\eqref{Spinfield} to the first order in $\boldsymbol L$ and substitute it in Eq.~\eqref{eq:kagomeDMI1}. The contribution proportional to $\boldsymbol m$ can be written in a compact form,
\begin{eqnarray}
\mathcal H_D^{(1)}=\boldsymbol{\mathcal B}\cdot\boldsymbol m ,
\end{eqnarray}
where $\boldsymbol{\mathcal B}$ is a ``magnetic field":
\begin{eqnarray}
\mathcal B_l=-i\text{tr}[\hat Z_{kl}\hat R^T\hat{\mathcal J}_k\hat R],
\end{eqnarray}
with $(Z_{kl})_{ab}=(S/6\hbar)\sum_{i,j}\sum_{c}D_{ij}^k(n^a_i\delta^{bc}-n_i^an_j^bn_j^c+\delta^{ac}n_j^b-n_i^an_j^bn_i^c)(\hat T^{-1}\hat R^T)_{cl}$.  Combing this term with Eqs.~\eqref{Lagrangian} and \eqref{KagomeDMI} and eliminating $\boldsymbol m$, the effective Lagrangian becomes:
\begin{eqnarray}\label{LagrangianwithDMI}
\mathcal L=\frac{\varrho}{2}(\boldsymbol{\Omega}_t-\boldsymbol{\mathcal B})^2-\mathcal U[\hat R],
\end{eqnarray}
where $\mathcal U[\hat R]=\text{tr}[\hat P_{ij}\partial_i\hat R^T\partial_j \hat R]+\mathcal H_D$.
The $\boldsymbol{\mathcal B}$ term breaks the rotation symmetry and gaps out the $\omega_1$ mode.

In a triangular lattice (see Fig.~\ref{fig_Symmetry}), the intrinsic DMI is forbidden by the lattice symmetry, while the ground state can be stabilized by the energy density,
\begin{eqnarray}\label{eq:anis}
\mathcal{H}_A&&=\sum_{i=1,2,3}-\mathcal  K(\hat{\boldsymbol n}_i\cdot\boldsymbol S_i)^2+\mathcal K_z(\hat{\boldsymbol z}\cdot\boldsymbol S_i)^2 ,
\end{eqnarray}
where $\mathcal K=K/A_{\text{uc}}$, $\mathcal K_z=K_z/A_{\text{uc}}$, with $K,K_z$ being the easy-axis and easy-plane anisotropy constants, respectively. By substituting $\boldsymbol S_i\approx \hat R\boldsymbol n_i$ in Eq~\eqref{eq:anis}, the anisotropy term gives a correction,
\begin{eqnarray}
\delta\mathcal{U}=\frac{\Delta}{2}(\theta_x^2+\theta_y^2)+3\mathcal KS^2\theta_z^2,
\end{eqnarray}
where $\Delta=3(\mathcal K_z+\mathcal K)S^2$.
When $\mathcal K\ll \mathcal K_z$, we can approximately neglect the easy-axis term, and thus the system approximately respects $U(1)$ symmetry. The Goldstone modes, $\omega_{2,3}$, acquire a gap, $\sqrt{\Delta/\varrho}$. A small gap in the $\omega_1$ mode is described by $\mathcal K$.

\textit{Spin superfluidity}---
In the following, we focus on the spin transport facilitated by approximate $U(1)$ symmetry. We assume that the system is driven by a weak perturbation when compared to the gap of the $\omega_{2}$ and $\omega_{3}$ modes, while large enough to overcome the barrier corresponding to the gap of the $\omega_{1}$ mode. By adding $\mathcal H_A$ or $\mathcal H_D$ to the Lagrangian \eqref{Sigmamodel} and neglecting the hard modes, the Lagrangian of the soft mode becomes:
\begin{eqnarray}\label{ULagrangian}\label{eq:softmode}
\mathcal L=\frac{\varrho}{2}\dot{\phi}^2-\frac{\mathcal A}{2}(\nabla\phi)^2,
\end{eqnarray}
where $\mathcal A=2(2\eta+\lambda+\mu)$. On the other hand, the third component of Eq.~\eqref{Totalspin} is reduced to $m_z\approx\varrho\partial_t\phi$.
Therefore, we arrive at a continuity equation,
\begin{eqnarray}
\partial_t m_z-\mathcal A\nabla^2 \phi=0,
\end{eqnarray}
where a spin current density with polarization along $z$-axis can be identified as
\begin{eqnarray}\label{eq:scurrent}
\boldsymbol j_s=-\mathcal A\boldsymbol\nabla\phi.
\end{eqnarray}

\begin{figure}
	 \includegraphics[width=1\linewidth]{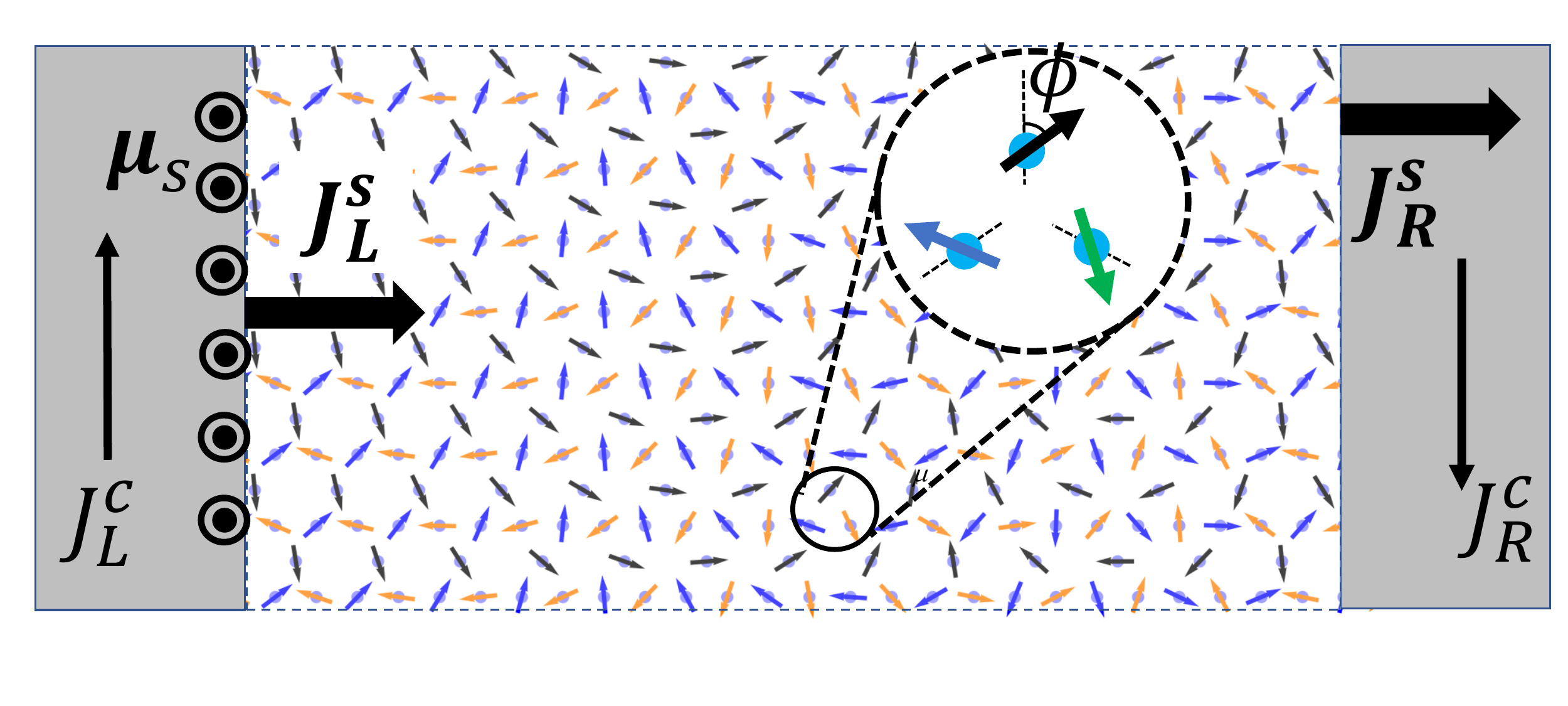}
 \caption{A nonlocal measurement setup containing normal metal/nAFM/normal heterostructure. A charge current in the left layer generates a spin accumulation $\boldsymbol\mu_s$ via spin Hall effect, which injects a spin current into nAFM layer. The spin current mediated by the collective modes in the middle layer passes the second interface by virtue of spin pumping effect. The pumped spin current is measured in the right layer via the inverse spin Hall effect. }\label{fig_setup}
\end{figure}

The spin superfluidity in nAFM will be affected by dissipation effects.
Within the Lagrangian formalism, we can add dissipation using the Rayleigh dissipation function~\cite{Gilbert2004}:
\begin{eqnarray}
\mathcal{R}=\frac{1}{2}Q_{ij}\dot{\boldsymbol{S}}_i\cdot\dot{\boldsymbol{S}}_j ,
\end{eqnarray}
where $\hat{Q}$ is a symmetric matrix with non-negative eigenvalues~\cite{Yuan_2019,PhysRevB.98.184402}. From symmetry considerations applied to three equivalent sublattices, we obtain that $Q_{ij}=r_1$ for $i=j$ and $Q_{ij}=r_2$ for $i\neq j$ where $r_{1}$ and $r_{2}$ are real parameters. The Rayleigh function can be approximately written as $\mathcal R=\text{tr}[\hat{\mathcal Q}\partial_t\hat R^T\partial_t \hat R]$ with $\mathcal Q_{ab}=\frac{1}{2}\sum_{ij}Q_{ij}n_i^an_j^b$ to describe dissipation of the soft mode, i.e.,
\begin{eqnarray}\label{Dissipation}
\mathcal{R}=\alpha s \dot{\phi}^2/2,
\end{eqnarray}
where $\alpha=3(r_1+2r_2)/s$ is a dimensionless dissipation parameter.

To activate spin dynamics in a magnetic insulator, a spin Hall current can be induced in a neighbouring normal metal layer (see Fig.~\ref{fig_setup}). A build-up of spin accumulation in a normal metal will then lead to injection of spin current into the magnetic insulator layer.  The boundary condition on the interface can be derived via the magnetoelectronic circuit theory~\cite{PhysRevLett.84.2481,BRATAAS2006157,PhysRevLett.117.207204,PhysRevB.96.100402,PhysRevB.101.224405}. When exchange interactions dominate, the spin injection and pumping together give the total spin current across the interface (see details in Supplemental Material):
\begin{eqnarray}\label{InterfaceSC}
\boldsymbol I_s=\frac{1}{4\pi}\hat{\mathcal G}^m\cdot (\boldsymbol{\mu}_s-\hbar\boldsymbol \omega).
\end{eqnarray}
where $\boldsymbol \omega$ is the instantaneous angular velocity for slow dynamics of the order parameter and $\hat{\mathcal G}^m$ is the generalized spin-mixing conductance tensor~\cite{PhysRevB.101.224405}. Under assumption of $C_3$ symmetry (or axial symmetry with respect to $\boldsymbol l$ axis), the tensor $\hat{\mathcal G}^m$ takes the following form:
\begin{eqnarray}\label{Spincondtensor}
\hat{\mathcal G}^m(\boldsymbol l)
&=&2\mathcal{G}^{\uparrow\downarrow}_r(\mathbbm{1}-\boldsymbol l\otimes\boldsymbol l)+2\mathcal{G}^{\uparrow\downarrow}_i(\boldsymbol l\times)+2\mathcal{G}_\parallel\boldsymbol l\otimes\boldsymbol l,\nonumber\\
\end{eqnarray}
where ${\cal G}^{\uparrow\downarrow}_{r(i)}= \sum_{mn}\operatorname{Re}(\operatorname{Im})(\delta_{nm}-r_{mn}^{\uparrow\uparrow}r_{mn}^{\downarrow\downarrow*})$, ${\cal G}_\parallel=\sum_{mn}(|r_{mn}^{\uparrow\downarrow}|^2+|r_{mn}^{\downarrow\uparrow}|^2)$, and $\boldsymbol l$ is the unit vector normal to the plane spanned by three-sublattice spins (cf. Eq.~(59) in Ref.~\cite{PhysRevB.101.224405}). Here,  $r^{\sigma\sigma^\prime}_{mn}$ stands for reflection amplitudes for electrons reflected from channel $n$ into channel $m$ in the normal metal and $\boldsymbol l$ is the quantization axis for $\sigma,\sigma^\prime=\uparrow,\downarrow$.

By writing the Euler-Lagrange equation with Rayleigh dissipation 
for Eqs.~\eqref{ULagrangian} and \eqref{Dissipation},
the dynamic equation for $\phi$ reads,
\begin{eqnarray}
\varrho\ddot{\phi}-\mathcal A\partial_x^2\phi+\alpha s\dot{\phi}=0.\label{eq:dynamic1}
\end{eqnarray}
We use a steady-state ansatz~\cite{PhysRevLett.112.227201, PhysRevB.90.094408}, $\phi(x,t)=\varphi(x)+\Omega t$,
where $\Omega$ is a constant frequency. For almost in-plane spin order, the angular velocity is $\boldsymbol\omega\approx \dot{\phi}\hat z$  and $\boldsymbol l=\hat z$. We also assume spin accumulation along $z$-direction, $\boldsymbol\mu_s=\mu\hat z$. Equations~\eqref{eq:scurrent}, \eqref{InterfaceSC}, and \eqref{Spincondtensor} then lead to the boundary conditions on the left ($x=0$) and right ($x=L$) interfaces: 
\begin{subequations}\label{Boundarycondition}
\begin{align}
-\mathcal A\partial_x\varphi(0)&=-\frac{g_L}{4\pi}(\hbar\Omega-\mu),\\
-\mathcal A\partial_x\varphi(L)&=\frac{g_R}{4\pi}\hbar\Omega,
\end{align}
\end{subequations}
where $g_a=\mathcal{G}_{\parallel,a}/\mathcal V$ ($a=L,R$) with $\mathcal V$ being the area of interface.
Combining the boundary condition and the steady-state ansatz, 
 we obtain,
 \begin{eqnarray}
 \Omega=\frac{\mu g_L}{\hbar(g_L+g_R+g_\alpha)},\quad
j^s_R=\frac{\mu}{4\pi}\frac{ g_Lg_R}{(g_L+g_R+g_\alpha)},
 \end{eqnarray}
where $g_\alpha=4\pi\alpha s L/\hbar$.

Above approximations need to be revisited when weak in-plane DMI or easy-axis anisotropy are present. We obtain a Lagrangian describing the soft mode:
\begin{eqnarray}\label{Lagrangian2}
\mathcal{L}[\phi]=
\frac{\varrho}{2}[(\partial_t\phi)^2-c_s^2(\nabla\phi)^2+m_s^2\cos2\phi],
\end{eqnarray}
where   $c_s=\sqrt{\mathcal A/\varrho}$ and $m_s$ is the mass term due to weak in-plane DMI or easy-axis anisotropy. To activate spin transport, the gradient needs to overcome the barrier induced by the gap, i.e., $|\nabla\phi|\geq m_s/c_s$.
The spiral-like  phase supporting spin superfluid will become energetically unstable 
when the field $\phi$ varies faster than $1/\xi$ where $\xi=\sqrt{\varrho c_s^2/\Delta}$ is  the characteristic length associated with the gap of $\omega_{2}$ and $\omega_{3}$ modes, i.e., $|\nabla\phi|\leq \xi^{-1}$. Thus, the spin-superfluid transport is only possible under the assumption,
\begin{eqnarray}\label{eq:cond}
m_s\sqrt{\varrho/\Delta}\ll 1.
\end{eqnarray}
We first discuss the kagome lattice nAFM with in-plane DMI in which case $m_s=\sqrt{3/2}D_p S/\hbar$. As estimated in Table.~\ref{table_material}, we find that different iron jarosites fulfil criteria \eqref{eq:cond} very well and hence are very promising for experimental realization of spin superfluidity.
In a triangular lattice, the easy-axis anisotropy hinders ideal spin superfluidity leading to the last term in Eq.~\eqref{Lagrangian2} with $m_s=3S\sqrt{K J}/\hbar$. Equation~\eqref{eq:cond} leads to the condition $\sqrt{K/K_z}\ll 1$.

The spin-superfluid transport can be measured in a nonlocal setup in Fig.~\ref{fig_setup}~\cite{PhysRevB.90.014428}. The spin Hall current builds up an effective spin accumulation, $\mu_s=(4\pi/g_L)J_{\text{SH}}^s$, where $J_{\text{SH}}^s=\vartheta_{\text{SH}}(\hbar/2e)J^c_L$ is the spin current induced by  the charge current $J^c_L$, and $\vartheta_{\text{SH}}$ is the spin Hall angle in the leads~\cite{RevModPhys.77.1375,PhysRevB.90.014428,PhysRevB.98.054424}. The spin current mediated by the collective dynamics of nAFM passes across the second interface by virtue of the spin pumping effect, and it is converted into a charge current in the right lead, $J^c_R=(\vartheta_{\text{SH}}\sigma/d)(\hbar/2e)\Omega$, where $\sigma$ and $d$ are, respectively, the conductivity and thickness of the right metal layer. The nonlocal transport is characterized by a drag coefficient, $\mathcal{D}=J_{R}^c/J^c_L=\mathcal{D}_0/(1+L/L_\alpha)$, where $\mathcal{D}_0=\pi\vartheta_{\text{SH}}^2\sigma\hbar/(2e^2gd)$, $g=g_L=g_R$, and $L_\alpha=\hbar g/(2\pi\alpha s)$. Assuming $\vartheta_{\text{SH}}=0.1$, $\sigma= 0.1 \, \mu\Omega^{-1}\cdot\text{cm}^{-1}$, $d= 1 \text{nm}$, $g\sim 10^{19}\text{m}^{-2}$, $\alpha= 10^{-3}$, $s\sim \hbar/a^3$, and a lattice constant $a\sim 1 \text{nm}$, we obtain $\mathcal{D}_0\sim 0.1$ and $L_\alpha\sim 1\mu \text{m}$. These results are similar to collinear systems \cite{PhysRevLett.112.227201, PhysRevB.90.094408} and show that the long crossover length $L_\alpha$ can be used as a key signature of spin superfluidity. 

\begin{table}[ht]
	\centering
	\begin{tabular}{|c| c| c| c| c|}
		\hline\hline 
		Material & $J$(meV) & $D_p/J$ & $D_z/J$&$m_s\sqrt{\varrho/\Delta}$ \\ [0.5ex] 
		\hline 
		KFe$_3$(OH)$_6$(SO$_4$)$_2$& 3.18& 0.062&-0.062&0.088\\
		AgFe$_3$(OH)$_6$(SO$_4$)$_2$& 3.18& 0.057&-0.053&0.088\\
		AgFe$_3$(OD)$_6$(SO$_4$)$_2$& 3.18& 0.075&-0.053&0.115
		\\[1ex]
		\hline 
	\end{tabular}
	\caption{Relevant material parameters for iron jarosites taken from Refs.~\cite{PhysRevLett.96.247201, PhysRevB.83.214406}.}
	 \label{table_material}
\end{table}

\textit{Conclusions}---
We have used an $SO(3)$-invariant field theory to describe  three-sublattice antiferromagnets with hexagonal lattice in an exchange interaction dominated limit. When weak interactions, such as DMI or anisotropy, are added, the symmetry is approximately reduced to $U(1)$. We have shown that in this limit, three-sublattice antiferromagnets can facilitate a spin superfluid transport.
Using generalized spin-mixing conductance, we have also described the injection of spin current and its power-law decay in a nonlocal experimental setup. Our results indicate that the magnitude of spin current is constrained by parasitic DMI or anisotropies, which can help in finding suitable materials. In particular, we estimate that iron jarosites can be promising for realizing spin superfluidity in noncollinear antiferromagnets. Noncollinear antiferromagnets hold promise for realizing spin flows with low dissipation and the theoretical framework presented here can be useful for exploring the interplay between transport phenomena~\cite{PhysRevB.92.220409,PhysRevB.99.180402} and topological defects, i.e., domain walls~\cite{PhysRevB.93.134429,PhysRevB.100.054415}, or skyrmions~\cite{PhysRevB.92.214439}. 

This work was supported by the U.S. Department of Energy, Office of Science, Basic Energy Sciences, under Award No. DE-SC0021019.

\bibliographystyle{apsrev}
\bibliography{SpinSuperfluid}

\newpage
\begin{center}
 \textbf{Supplemental Material}  
\end{center}

\section{DMI or anisotropy induced gap}

In the main text, we have shown that the out-of-plane DMI or easy-plane anisotropy can suppress spin rotations except for the one with respect to $z$-direction, thus reduce the $SO(3)$ symmetry to a $U(1)$-rotation symmetry. Here, we show that in the presence of a slowly varying $\phi$-angle spiral background allowed by the $U(1)$-rotation symmetry, the local spectrum of hard rotation modes with respect to $x$- or $y$-directions are always separated from the soft rotation by a constant gap. To capture this scenario, we use the rotation matrix $\hat R=\exp [-i\boldsymbol\theta \cdot\hat{\boldsymbol{\mathcal J}}]\exp[-i\phi \hat{\mathcal J}_z]$, where the first component describes the spin waves ($|\boldsymbol\theta|\ll 1$), the second component parametrizes the local background state. With this representation, the local spin-wave Lagrangian for a system with out-of-plane DMI or easy-plane anisotropy  reads
\begin{eqnarray}
\mathcal L=\frac{\varrho}{2}\dot{\boldsymbol\theta}^2-\mathcal U[\phi]
\end{eqnarray}
with
\begin{eqnarray}
\mathcal U[\phi]=&&\frac{\mathcal A}{2}(\nabla\theta_z)^2+\frac{\Delta}{2}(\theta_x^2+\theta_y^2)\nonumber\\
&&+(\mathcal C_1+\mathcal C_2\sin^2\phi)[(\partial_x\theta_x)^2+(\partial_y\theta_y)^2]\nonumber\\
&&+(\mathcal C_1+\mathcal C_2\cos^2\phi)[(\partial_x\theta_y)^2+(\partial_y\theta_x)^2]\nonumber\\
&&+C_2\sin(2\phi)[\partial_y\theta_x\partial_y\theta_y-\partial_x\theta_x\partial_x\theta_y]\nonumber\\
&&+2(\mu\sin^2\phi-\lambda\cos^2\phi)\partial_x\theta_x\partial_y\theta_y\nonumber\\
&&+2(\lambda\sin^2\phi-\mu\cos^2\phi)\partial_y\theta_x\partial_x\theta_y.
\end{eqnarray}
Here, $\mathcal C_1=\kappa+\eta$, $\mathcal C_2=\lambda+\mu$, $\Delta$ is a mass term induced by the out-of-plane DMI or easy-plane anisotropy. Specifically, $\Delta=-3\sqrt{3}\mathcal D_zS^2$ for the kagome nAFM with an out-of-plane DMI and $\Delta=3\mathcal K_zS^2$ for the triangle nAFM with an easy-plane anisotropy. By applying the Euler-Lagrange equation and converting the equations to frequency-momentum space, we obtain 
\begin{widetext}
\begin{eqnarray}
&&\varrho\omega^2\theta_z=\mathcal A k^2\theta_z,\nonumber\\
&&\varrho\omega^2\left(
\begin{array}{cc}
     \theta_x \\
     \theta_y 
\end{array}\right)=\left(
\begin{array}{cc}
     \Delta+(2\mathcal C_1+\mathcal C_2) k^2+\mathcal C_2\cos(2\phi)(k_y^2-k_x^2)& \mathcal C_2\sin(2\phi)(k_y^2-k_x^2)-2\mathcal C_2\cos(2\phi) k_xk_y\\
      \mathcal C_2\sin(2\phi)(k_y^2-k_x^2)-2\mathcal C_2\cos(2\phi) k_xk_y& \Delta+(2\mathcal C_1+\mathcal C_2) k^2-\mathcal C_2\cos(2\phi)(k_y^2-k_x^2)
\end{array}\right)
\left(
\begin{array}{cc}
     \theta_x \\
     \theta_y 
\end{array}\right).\nonumber\\
\end{eqnarray}
The spectrum is solved from equations above
\begin{eqnarray}
\omega_1=\sqrt{\frac{\mathcal A}{\varrho}}k,\qquad
\omega_{2,3}=\Bigg(\frac{\Delta}{\varrho}+\frac{k^2}{\varrho}\big[2\mathcal C_1+\mathcal C_2 \pm\mathcal C_2\sqrt{\cos^2(2\phi)\cos^2(2\xi_{\boldsymbol k})+\sin^2(2\phi+2\xi_{\boldsymbol k})}\big]\Bigg)^{1/2},
\end{eqnarray}
\end{widetext}
where $(k_x,k_y)=k(\cos\xi_{\boldsymbol k}, \sin\xi_{\boldsymbol k})$. We found that the size of the gap induced by the mass term is independent of the local background. In the limit $\phi\rightarrow 0$, the Lagrangian describes the spin wave in the ground state and the spectrum recovers the three Goldstone modes in the absence of DMI or anisotropy, i.e., $\Delta=0$. When the system is driven by a weak perturbation with a characteristic energy scale below the gap induced by the mass term, the dynamics and spatial variation of hard modes are unactivated. In this regime, we can concentrate on the slow, long-wavelength dynamics of the soft $\phi$-mode by dropping all variations of hard modes, as described by Eq.~\eqref{eq:softmode}.

\section{Boundary condition}\label{Ap_bondarycondition}

According to the magnetoelectronic circuit theory, the general boundary current across a magnet/normal metal interface on the metal side is~\cite{PhysRevLett.84.2481,BRATAAS2006157}
\begin{eqnarray}\label{Circuittheory}
 \hat i_N=\frac{e}{h}\sum_{mn}\left[\hat t^\prime_{mn}\hat f^M(\hat t^\prime_{mn})^\dagger-\left(\delta_{mn}\hat f^N-\hat r_{mn}\hat f^N(\hat r_{mn})^\dagger\right)\right].\nonumber\\
\end{eqnarray}
Here, $\hat f^M,\hat f^N$ are distribution function matrix for magnet and normal metal, respectively, $\hat t^\prime_{mn}$ donates the spin-dependent transmission amplitude for electrons transmitted from channel $n$ in node magnet into channel $m$ in normal metal, $\hat r_{mn}$ stands for the spin dependent reflection  amplitude in the normal metal from channel $m$ to channel $n$. In our discussion, the transmission matrix vanishes, $\hat t^\prime_{mn}=0$, as we consider an insulating magnet. The current matrix and distribution function matrix can be decomposed as below~\cite{RevModPhys.77.1375} 
\begin{eqnarray}\label{Decomposition}
 &&\hat i_N=\frac{1}{2}(\hat{\sigma}_0i^N_c+\frac{e}{\hbar}\hat{\boldsymbol{\sigma}}\cdot\boldsymbol i^N_s),\\
 &&\hat f^a=\hat{\sigma}_0f^a_0+\hat{\boldsymbol{\sigma}}\cdot \boldsymbol f^a_s,\quad (a= M,N).
\end{eqnarray}
Here, $i^N_c$ and $\boldsymbol i^N_s$ stand for the charge and spin current at a given energy, respectively; $f_0^a$ reflects the charge accumulation, $\boldsymbol f^a_s$ is a vector whose direction depends on the device configuration and applied biases; $\sigma_0$ is the identity matrix and $\boldsymbol \sigma=\{\sigma_x,\sigma_y,\sigma_z\}$ with $\sigma_i$ being Pauli matrix.

 With applying spin bias in the normal metal, a spin accumulation, $\boldsymbol\mu_s$, can be built up near the interface providing the metal layer is a poor spin sink. This spin accumulation can inject a spin current across the normal metal/magnetic insulator interface. At a given energy $\epsilon$, the spin current on the magnetic insulator side (opposite to the metal side) is obtained from the  Eq.~\eqref{Circuittheory} and Eq.~\eqref{Decomposition} ~\cite{PhysRevLett.117.207204, PhysRevB.101.224405}
\begin{eqnarray}\label{Spincurrent}
\boldsymbol i_s(\epsilon)=-\boldsymbol i^N_s(\epsilon)=\frac{1}{2\pi}\hat{\mathcal G}^m\cdot\boldsymbol f_s(\epsilon),
\end{eqnarray}
where $\hat{\mathcal G}^m$ is a spin conductance tensor. The spin accumulation vector and distribution function are connected as below
\begin{eqnarray}\label{Spinaccumulation}
 \boldsymbol \mu_s=\int d\epsilon \text{tr}[\hat{\boldsymbol \sigma}\hat f(\epsilon)]=2\int d\epsilon\boldsymbol f_s(\epsilon).
\end{eqnarray}
By integrating Eq.~\eqref{Spincurrent} over energy and using Eq.~\eqref{Spinaccumulation}, the total spin current across the interface is obtained
\begin{eqnarray}\label{Spincurrent1}
\boldsymbol I_s=\frac{1}{4\pi}\hat{\mathcal G}^m\cdot\boldsymbol\mu_s,
\end{eqnarray}
In our discussion, we dropped the contribution form transmission across the interface as we mentioned in the beginning. The tensor $\hat{\mathcal G}^m$ is solely expressed in terms of reflection matrix elements~\cite{PhysRevLett.117.207204, PhysRevB.101.224405}
\begin{eqnarray}\label{Spinconductance}
  \mathcal G^m_{ij}=2\delta^{kl}_{ij}\sum_{mn}\mathcal R_{mn}^{kl}+\mathcal R_{mn}^{lk}+i\epsilon_{kl\nu}(\mathcal R_{mn}^{0\nu}-\mathcal R_{mn}^{\nu0}),
\end{eqnarray}
where
\begin{eqnarray}
 \mathcal R_{mn}^{\mu\nu}=\frac{1}{4}\text{Tr}[(\hat r_{mn}\otimes\hat r^\ast_{mn})\cdot(\sigma^\mu\otimes\sigma^\nu)],
\end{eqnarray}
and $\delta^{kl}_{ij}=\delta_{ij}\delta_{kl}-\delta_{ik}\delta_{jl}$. The explicit form of $\hat{\mathcal G}^m$ is
\begin{widetext}
\begin{eqnarray}\label{Spinconductance1}
\hat{\mathcal G}^m=\left(
\begin{array}{c c c}
    |r^{\downarrow\uparrow}-r^{\uparrow\downarrow}|^2+|r^{\downarrow\downarrow}-r^{\uparrow\uparrow}|^2 &2\text{Im}(r^{\downarrow\uparrow\ast}r^{\uparrow\downarrow}+r^{\downarrow\downarrow}r^{\uparrow\uparrow\ast}) &2\text{Re}(r^{\downarrow\downarrow\ast}r^{\uparrow\downarrow}-r^{\downarrow\uparrow\ast}r^{\uparrow\uparrow}) \\
    2\text{Im}(r^{\downarrow\uparrow\ast}r^{\uparrow\downarrow}+r^{\downarrow\downarrow\ast}r^{\uparrow\uparrow}) &  |r^{\downarrow\uparrow}+r^{\uparrow\downarrow}|^2+|r^{\downarrow\downarrow}-r^{\uparrow\uparrow}|^2& 2\text{Im}(r^{\downarrow\downarrow}r^{\uparrow\downarrow\ast}+r^{\downarrow\uparrow^\ast}r^{\uparrow\uparrow})\\
    2\text{Re}(r^{\downarrow\downarrow\ast}r^{\downarrow\uparrow}-r^{\uparrow\downarrow\ast}r^{\uparrow\uparrow})&2\text{Im}(r^{\downarrow\downarrow\ast}r^{\downarrow\uparrow}+r^{\downarrow\uparrow}r^{\uparrow\uparrow\ast})&2(|r^{\downarrow\uparrow}|^2+|r^{\uparrow\downarrow}|^2)
\end{array}\right).
\end{eqnarray}
\end{widetext}
 Assume the coupling between normal metal and magnet on the boundary is dominated by exchange interaction, thus the boundary condition stays unchanged upon a $SO(3)$ rotation in the spin space. Therefore, we are able to consider a special case with all three-sublattice spins lying in the plane of the interface and chose the direction normal to the interface as $z$-direction. The system should respect a rotation symmetry with respect to $z$-axis.

The tensor in principle should be a function of magnetic spin orders on the boundary, i.e., $\hat{\mathcal G}^m=\hat{\mathcal G}^m(\{\boldsymbol S_i\})$. For a given symmetry $\hat{\mathcal O}$ of the system, the tensor respects the condition
\begin{eqnarray}
\hat{\mathcal G}^m(\{\det[\hat{\mathcal O}]\hat{\mathcal O}\boldsymbol S_i\})=\hat{\mathcal O}\hat{\mathcal G}^m(\{\boldsymbol S_i\})\hat{\mathcal O}^{-1}.
\end{eqnarray}
When the symmetry operation is unitary ($\det[\hat{\mathcal O}]=1$), the variation in spin space does not change the tensor form as a result of the $SO(3)$ symmetry, i.e., $\hat{\mathcal G}^m(\{\det[\hat{\mathcal O}]\hat{\mathcal O}\boldsymbol S_i\})=\hat{\mathcal G}^m(\{\boldsymbol S_i\})$. Therefore, the unitary symmetry constraint  on the tensor reads
\begin{eqnarray}
\hat{\mathcal G}^m(\{\boldsymbol S_i\})=\hat{\mathcal O}\hat{\mathcal G}^m(\{\boldsymbol S_i\})\hat{\mathcal O}^{-1}.
\end{eqnarray}
 Applying the rotation symmetry with respect to $z$-direction produces
\begin{eqnarray}
\hat{\mathcal G}^m_0=\left(
\begin{array}{ccc}
     x_1&x_2&0  \\
     -x_2&x_1&0\\
     0&0&x_3
\end{array}
\right),
\end{eqnarray}
where $x_1,x_2,x_3$ are coefficients that will be determined later.
Since the system has a $SO(3)$ symmetry, we are allowed to rotate the perpendicular direction of the spin-spanned plane (spin quantization axis) from $z$-direction to arbitrary direction $\boldsymbol l=(\sin\theta\cos\phi,\sin\theta\sin\phi,\cos\theta)$ by appling a rotation matrix $\hat{\mathcal O}(\boldsymbol l)=\text{exp}[-i\hat{\mathcal J}_z\phi]\text{exp}[-i\hat{\mathcal J}_y\theta]$. The spin conductance tensor under a representation with new quantization axis becomes
\begin{eqnarray}\label{spin-mixing}
\hat{\mathcal G}^m_0\rightarrow \hat{\mathcal G}^m(\boldsymbol l)= \hat{\mathcal O}(\boldsymbol l)\hat{\mathcal G}^m_0\hat{\mathcal  O}^T(\boldsymbol l),
\end{eqnarray}
where $\hat{\mathcal G}^m(\boldsymbol l)$ is explicitly written as
\begin{eqnarray}\label{Spinconductancesymmetry}
\hat{\mathcal G}^m(\boldsymbol l)&=&\left(
\begin{array}{ccc}
     x_1&0&0  \\
     0&x_1&0\\
     0&0&x_1
\end{array}
\right)
-x_2\left(
\begin{array}{ccc}
     0&-l_z&l_y  \\
     l_z&0&-l_x\\
     -l_y&l_x&0
\end{array}
\right)\nonumber\\
&&+(x_3-x_1)\left(
\begin{array}{ccc}
     l_x^2&l_xl_y&l_xl_z  \\
     l_yl_x&l^2_y&l_yl_z\\
     l_zl_x&l_zl_y&l_z^2
\end{array}
\right)\nonumber\\
&=&x_1(\mathbbm{1}-\boldsymbol l\otimes\boldsymbol l)-x_2(\boldsymbol l\times)+x_3\boldsymbol l\otimes\boldsymbol l.
\end{eqnarray}

Now, we can fix the coefficient in Eq.~\eqref{Spinconductancesymmetry} by comparing it with Eq.~\eqref{Spinconductance1}. In the frame with $\boldsymbol l$ being the spin quantization axis, only the elements fitting the form of Eq.~\eqref{Spinconductancesymmetry} take finite values, all others are forced to vanish by symmetry. Combing with the scattering matrix normalization relation, $\sum_{mn}\hat r_{mn}\hat r_{mn}^\dagger=M\sigma_0$, the tensor takes the form in the main text, i.e., $x_1=2\mathcal G^{\uparrow\downarrow}_r$, $x_2=-2\mathcal G^{\uparrow\downarrow}_i$, and $x_3=2\mathcal G_\parallel$.

Above, we discussed spin current across the boundary induced by spin accumulation. However, when magnetic spins start to process, the dynamic variation of spins will also generate a pumping current.
In a steady state with adjacent metal and magnet staying in a (dynamic) mutual equilibrium, if one observes the system in a rotation frame stick-on the magnet, the net spin current will vanish as spins in the magnet are static. This could be understood as that the spin accumulation is canceled  by a rotation induced effective magnetic field $\boldsymbol H_{\text{eff}}=\hbar\boldsymbol{\omega}$ with $\boldsymbol \omega$ being the vectorial angular velocity \cite{RevModPhys.77.1375,PhysRevB.66.224403}. This means that the total current in the laboratory frame should take the form 
\begin{eqnarray}\label{boundaryspincurrent}
\boldsymbol I_s=\frac{1}{4\pi}\hat{\mathcal G}^m\cdot (\boldsymbol{\mu}_s-\hbar\boldsymbol \omega).
\end{eqnarray}

In the trilayer structure discussed in the main text, the boundary current density across two interfaces from left to right are
\begin{eqnarray}
&&\boldsymbol j_s|_L=-\frac{\hat g_L}{4\pi}\cdot (\hbar\boldsymbol\omega-\boldsymbol \mu_s),\nonumber\\
&&\boldsymbol j_s|_R=\frac{\hat g_R}{4\pi}\cdot\hbar\boldsymbol\omega,
\end{eqnarray}
where we assumed that the spin accumulation in the right normal metal is zero. Here,
\begin{eqnarray}
\hat g_a=\hat{\mathcal G}^m_a/\mathcal V,\qquad a=L,R,
\end{eqnarray}
with $\mathcal V$ being the area of the interface. When  all spins lie in the plane, both vectorial angular velocity and spin accumulation are pointing along the out-of-plane direction, the boundary condition will be reduced to  Eq.~\eqref{Boundarycondition}.

\section{Spin pumping}
We discussed the spin pumping effect by switching to a frame rotating with the magnet in the last section. Here, we apply a parametric pumping theory~\cite{PhysRevB.58.R10135} to confirm this argument. 
Assuming the scattering matrix depends on time through a set of real parameter $\{X_i(t)\}$, the current pumped to the normal metal is~\cite{7bcfbe1bf0bb4851982da0b2b2c65ae7}
\begin{eqnarray}\label{currentmatrix}
\hat{I}_N=e\frac{\partial \hat n}{\partial X_j}\frac{dX_j}{dt}
\end{eqnarray}
where 
\begin{eqnarray}
\frac{\partial \hat n}{\partial X_j}=\frac{1}{4\pi i}\sum_{mn}\frac{\partial \hat r_{mn}}{\partial X_j}\hat r_{mn}^\dagger+\text{H.c.}
\end{eqnarray}
Here, the scattering matrix is reduced to the reflection matrix in the normal metal because the transmission amplitude from the magnet to the normal metal is zero. The reflection matrix follows the rotation of order parameters by
\begin{eqnarray}
\hat r_{mn}(t)=U\hat r^{(0)}_{mn} U^\dagger
\end{eqnarray}
where $r^{(0)}_{mn}$ is the reflection matrix for the initial state, $U$ is a $SU(2)$ spin rotation matrix corresponding to the $SO(3)$ rotation of the order parameters. We use the representation $U=\exp[i\beta(\boldsymbol\nu\cdot\hat{\boldsymbol\sigma})/2]$ with $\boldsymbol\nu$ and $\beta$ being the instantaneous rotation axis and angle, respectively. At a given moment, $\hat I_N=e(\partial\hat n/\partial\beta) \dot{\beta}$, namely the rotation axis is instantaneously static. Therefore, we have
\begin{eqnarray}\label{particlematrix}
\frac{\partial\hat n}{\partial \beta}=&&\frac{1}{4\pi i}\sum_{mn}\frac{\partial\hat r_{mn}}{\partial\beta}\hat r_{mn}^\dagger+\text{H.c.}\nonumber\\
=&&\frac{1}{4\pi}\sum_{mn}[(\boldsymbol\nu\cdot\hat{\boldsymbol\sigma})\delta_{mn}-\hat r_{mn}(\boldsymbol\nu\cdot\hat{\boldsymbol\sigma})\hat r^\dagger_{mn}],
\end{eqnarray}
where we used $\partial_\beta U=i(\boldsymbol\nu\cdot\hat{\boldsymbol\sigma}/2) U$ and $\sum_{mn}\hat r_{mn}\hat r_{mn}^\dagger=M\sigma_0$ with $M$ being the number of channels in the normal metal. The current matrix can be decomposed as $\hat I=\frac{1}{2}(\hat\sigma_0 I_c^N+\frac{e}{\hbar}\hat{\boldsymbol\sigma}\cdot\boldsymbol I^N_s)$. By plugging Eq.~\eqref{particlematrix} into Eq.~\eqref{currentmatrix} and taking $I_{s,i}=\frac{\hbar}{e}\text{Tr}[\hat\sigma_i\hat I_N]$, we obtain the pumped spin  current in the metal layer
\begin{eqnarray}
I_{s,i}=\frac{\hbar}{4\pi}\sum_{mn}\big(2\delta_{ij}\delta_{mn}-\text{Tr}[\sigma_i\hat r_{mn}\hat\sigma_j\hat r^\dagger_{mn}]\big)\omega_j,
\end{eqnarray}
where $\omega_j=\dot{\beta}\nu_j$ is the instantaneous angular velocity. It can be shown that the tensor in the equation above agrees with spin-mixing conductance  Eq.~\eqref{Spinconductance1}. By fitting the above result to Eq.~\eqref{Spinconductancesymmetry}, we can verify that the spin pumping expression is consistent with Eq.~\eqref{boundaryspincurrent}.

\end{document}